\documentclass[final,3p,times,twocolumn]{elsarticle}
\biboptions{comma,sort&compress}
\usepackage{ecrc}
\usepackage{here}
\usepackage{graphicx}
\usepackage{epsfig}
\usepackage{epstopdf}
\usepackage{amsmath}

\def\nin{\noindent}
\def\beq{\begin{equation}}
\def\eeq{\end{equation}}
\def\bea{\begin{eqnarray}}
\def\eea{\end{eqnarray}}
\def\nnb{\nonumber}
\def\la{\langle}
\def\ra{\rangle}

\def\lrar{\Longrightarrow}

\usepackage{graphicx}
\usepackage{here}
\def\beq{\begin{equation}}
\def\eeq{\end{equation}}
\def\bea{\begin{eqnarray}}
\def\eea{\end{eqnarray}}
\def\bq{\begin{quote}}
\def\eq{\end{quote}}
\def\ve{\vert}
\def\nnb{\nonumber}

\def\lrar{\Longrightarrow}
		
\def\nnb{\nonumber}
\def\la{\langle}
\def\ra{\rangle}
\def\nin{\noindent}
\def\ba{\begin{array}}
\def\ea{\end{array}}

\def\als{\alpha_s}

\def\gg2{ \la\alpha_s G^2 \ra}
\def\gg3{g^3f_{abc}\la G^aG^bG^c \ra}
\def\ggg4{\la\als^2G^4\ra}

\def\beq{\begin{equation}}
\def\enq{\end{equation}}
\def\beqa{\begin{eqnarray}}
\def\enqa{\end{eqnarray}}
\def\nnb{\nonumber}

\def\MeV{\nobreak\,\mbox{MeV}}

\def\keV{\nobreak\,\mbox{keV}}

\newcommand{\rag}{\rangle}
\newcommand{\lag}{\langle}

\def\gg{\lag g^{2}_{s} G^2 \rag}
\def\ggg{\lag g^{3}_{s}G^3\rag}

\volume{00}
\firstpage{1}
\journalname{Nuclear and Particle Physics Proceedings }
\runauth{D. Rabetiarivony}
\jnltitlelogo{Nuclear and Particle Physics Proceedings }

\begin{document}
\begin{frontmatter}

\title{$0^+$ XTZ states from QCD spectral sum rules\tnoteref{text1}}

\author[label1]{R. M. Albuquerque}
\ead{raphael.albuquerque@uerj.br}
\address[label1]{Faculty of Technology,Rio de Janeiro State University (FAT,UERJ), Brazil}

\author[label2,label3]{S. Narison}
\ead{snarison@yahoo.fr}
\address[label2]{Laboratoire
Univers et Particules de Montpellier (LUPM), CNRS-IN2P3, Case 070, Place Eug\`ene
Bataillon, 34095 - Montpellier, France}
\address[label3]{Institute of High-Energy Physics of Madagascar (iHEPMAD), University of Ankatso, Antananarivo 101, Madagascar}

\author[label3]{D. Rabetiarivony\fnref{fn2}}
\fntext[fn2]{Speaker}
\ead{rd.bidds@gmail.com} 


\tnotetext[text1]{Talk given at QCD22 International Conference (4--7 July 2022, Montpellier--FR)}

\pagestyle{myheadings}
\markright{ }

\begin{abstract}
\noindent
We review our results in\,\cite{ANR22} for the masses and couplings of $T_{ccqq'}\, (J^P=0^+)$ states from (inverse) QCD Laplace sum rule (LSR), their ratios ${\cal R}$ and double ratio of sum rules (DRSR) within stability criteria and including Factorized Next-to-Leading Order (FNLO) Perturbative (PT) corrections and  Lowest Order (LO) QCD condensates up to $\lag G^3 \rag$. We show that combining ${\cal R}$ and DRSR can provide more precise results. Calibrated to the observed $X_c(3872)$ and $T^{1^+}_{cc}(3875)$, ${\cal R}$ combined with DRSR lead to a more  precise prediction of $M_{T^{0^+}_{cc}}=3883(3)\MeV$. In a similar way, calibrated to the new prediction of $T^{0^+}_{cc}$ ${\cal R} \oplus$DRSR lead to the improved mass predictions: $M_{T^{0^+}_{cc\bar{s}\bar{u}}}=3927(6)\MeV$ and $M_{T^{0^+}_{cc\bar{s}\bar{s}}}=3993(11)\MeV$. We extend our analysis to the bottom sector and compare our results with the ones from different LSR predictions and some other determinations (lattice, quark and potential models,...) in the literature.
\end{abstract} 
\scriptsize
\begin{keyword}
QCD Spectral Sum Rules \sep Perturbative and Non-perturbative QCD \sep Exotic hadrons \sep Masses and Decay constants.
\end{keyword}
\end{frontmatter}
\section{Introduction}
\nin In a series of papers \cite{ANR21,ANRR1,ANRR1a,QQQQ,ANRR2,ANR1,NR1,AFNR,SNX1,SNX2,SU3}, we have used the inverse Laplace Transform (LSR) \cite{BELL,BELLa,BNR,BERT,NEUF,SNR} of QCD spectral sum rules (QSSR)\footnote{For reviews, see \cite{SVZa,Za,SNB1,SNB2,SNB3,CK,YND,PAS,RRY,IOFF,DOSCH}} to extract the masses and couplings of some heavy-light and fully heavy molecules and tetraquark states.

\nin In this talk, based on the paper in Ref.\,\cite{ANR22}, we improve the existing QSSR results for the masses and couplings of $T_{cc\bar u\bar d}$ ($0^+$) by combining the direct mass determination from the ratios ${\cal R}$ of LSR with the ratio of masses from DRSR. In so doing, we will use the $X_c(3872)$\,\cite{X3872-EXP} and $T^{1^+}_{cc}(3875)$\,\cite{LHCb} as inputs in our DRSR approach. We pursue our analysis by studying their SU(3) partners: $T_{cc\bar{u}\bar{s}}$ and $T_{cc\bar{s}\bar{s}}$ states. Finally, we complete our analysis by extending it to the bottom sector ($T_{bbqq'}$).
\section{The Laplace sum rule}
\nin We shall work with the finite energy version of the QCD inverse Laplace sum rules and their ratios:
\bea
\hspace*{-0.2cm} {\cal L}^Q_n\vert_{\cal H}(\tau,\mu)&=&\int_{t^{2}_{0}}^{t_c} dt~t^n~e^{-t\tau}\frac{1}{\pi} \mbox{Im}~\Pi^{(0)}_{\cal H}(t,\mu)~;\nnb \\
 {\cal R}^c_{\cal H}(\tau)&=&\frac{{\cal L}^Q_{n+1}\vert_{\cal H}} {{\cal L}^Q_n\vert_{\cal H}},
\label{eq:lsr}
\eea
where $t_0=2M_Q+m_q+m_{q'}$ is the hadronic threshold, $M_Q$ ($Q\equiv c,b$) and $m_q,\,m_{q^{'}}$ ($q,q'\equiv u,d,s$) are respectively the heavy and light quarks masses (we shall neglect the $u,d$ quark masses), $\tau$ is the LSR variable, $n=0,\cdots$ is the degree of moments and $t_c$ is the "QCD continuum" threshold.\\
From the ratio of LSR, one can extract the mass squared at the optimization point $\tau_0$\,:
\beq
 {\cal R}^c_{\cal H}(\tau_0)= M_{\cal H}^2.
\eeq
We shall also work with the double ratio of sum rules (DRSR)\,\cite{DRSR88}\,:
\beq
r_{{\cal H'}/{\cal H}}(\tau_0)\equiv \sqrt{\frac{{\cal R}^c_{\cal H'}}{{\cal R}^c_{\cal H}}}=\frac{M_{\cal H'}}{M_{\cal H}},
\eeq
which can be free from systematics provided that ${\cal R}^c_{\cal H}$ and ${\cal R}^c_{\cal H'}$ optimize at the same values of $\tau$ and of $t_c$:
\beq
 \tau_0\vert_{\cal H}\simeq \tau_0\vert_{\cal H'}~,~~~~~~~~~~~~t_c\vert_{\cal H}\simeq t_c\vert_{\cal H'}~.
\eeq
The spectral function $\mbox{Im}~\Pi^{(0)}_{\cal H}(t,\mu)$ associated to the scalar interpolating operators can be evaluated from the two-point correlator:
\bea
\hspace*{-0.6cm}
\Pi^{\mu\nu}_{\mathcal{H}}(q^2)\hspace*{-0.2cm}&=& \hspace*{-0.2cm} i \int \hspace*{-0.1cm} d^4 x\, e^{-i q x}\lag 0 \ve \mathcal{T} \mathcal{O}^{\mu}_{\mathcal{H}}(x)(\mathcal{O}^{\nu}_{\mathcal{H}}(x))^{\dag} \ve 0 \rag,\nnb \\
&\equiv &\hspace*{-0.2cm} -\left( g^{\mu\nu}-\frac{q^{\mu}q^{\nu}}{q^2}\right)\Pi^{(1)}_{\mathcal{H}}(q^2)+\frac{q^{\mu}q^{\nu}}{q^2}\Pi^{(0)}_{\mathcal{H}}(q^2),
\eea
where $\mathcal{O}^{\mu}_{\mathcal{H}}(x)$ are the interpolating currents describing the four-quark states.

\nin The QCD expressions of the leading order (LO) spectral functions up to dimension 6 condensates, details about the FNLO PT corrections and the different QCD input parameters are given in Ref.\,\cite{ANR22,PICH,NPIV,HAGIWARA,RRY}.
\vspace*{-0.3cm}
\subsection{Interpolating currents}
\nin
We shall be concerned with the $J^P=0^+$ interpolating currents, the $X_c(1^+)$ and $T_{cc}(1^+)$ states given in Table \ref{tab:current}.
{\small
\begin{table*}[hbt]
\setlength{\tabcolsep}{2.5pc}
{\small
\begin{tabular*}{\textwidth}{@{}lll@{\extracolsep{\fill}}l}
&\\
\hline
\hline
States&$I(J^P)$& $\bar 3_c3_c$ Four-quark  Currents\\
\hline
\\
$Z_c$ &$(1^{+})$&$ {\cal O}_{A_{cq}}=\epsilon_{ijk}\epsilon_{mnk}\big{[}(q^T_i\,C\gamma_5\,c_j)(\bar q'_m \gamma_\mu C\, \bar c_n^T)\,+b,(q^T_i\,C\,c_j)(\bar q'_m\gamma_\mu \gamma_5 C\,\bar c_n^T)\big{]}$\\
&&$ {\cal O}_{D^*_qD_q}=(\bar c\gamma_\mu q)(\bar q'\,i\gamma_5c)$  \\
\\
$X_c $&$(1^{+})$
&$ {\cal O}^3_{X} = \epsilon_{i j k} \:\epsilon_{m n k} \big{[}\left(
  q_i^T\, C \gamma_5 \,c_j \right) \left( \bar{c}_m\, \gamma^\mu
  C \,\bar{q}_n^T\right) + \left(
  q_i^T\, C \gamma^\mu \,c_j \right) \left( \bar{c}_m\, \gamma_5
  C \,\bar{q}_n^T\right)\big{]}$\\
&& $ {\cal O}^6_{X} = \epsilon_{i j k} \:\epsilon_{m n k}\big{[} \left(
  q_i^T\, C \gamma_5 \lambda_{ij}^a\,c_j \right) \left( \bar{c}_m\, \gamma^\mu
  C\lambda_{mn}^a \,\bar{q}_n^T\right) +  \left(
  q_i^T\, C \gamma^\mu \lambda_{ij}^a\,c_j \right) \left( \bar{c}_m\, \gamma_5
  C \lambda_{mn}^a\,\bar{q}_n^T\right)\big{]}$ \\
&& $ {\cal O}_{D^*_qD_q}= \frac{1}{\sqrt{2}}\big{[}(\bar q\gamma_5 c) (\bar c\gamma_\mu q) -  (\bar q\gamma_\mu c) (\bar c\gamma_5 q)\big{]} $ \\
&& $ {\cal O}_{\psi\pi} = (\bar c\gamma_\mu\lambda^a c)(\bar q\gamma_5\lambda^a q)$\\
\\
$T_{cc\bar u\bar d}$&$0(1^{+})$
 &  $ {\cal O}_T^{1^+} = \frac{1}{\sqrt{2}}\epsilon_{i j k} \:\epsilon_{m n k} \left(
 c_i^T\, C \gamma^\mu \,c_j \right) \big{[} \left( \bar{u}_m\, \gamma_5
 C \,\bar{d}_n^T\right) -  \left( \bar{d}_m\, \gamma_5
 C \,\bar{u}_n^T\right)\big{]}$ \\
 $T_{cc\bar u\bar s}$&$\frac{1}{2}(1^{+})$
 & $ {\cal O}_{T^{1^+}_{us}}   = \epsilon_{i j k} \:\epsilon_{m n k}
    \left( c_i \, C \gamma^{\mu } c_j^T \right) 
    \left( \bar{u}_m \,\gamma_5 C \bar{s}_n^T \right)$\\
  \\
$T_{cc\bar u\bar d}$&$1(0^{+})$
&$ {\cal O}_T^{0^+} = \frac{1}{\sqrt{2}}\epsilon_{i j k} \:\epsilon_{m n k} \left(c_i^T\, C \gamma^\mu \,c_j \right) \big{[} \left( \bar{u}_m\, \gamma_\mu C \,\bar{d}_n^T\right) + \left( \bar{d}_m\, \gamma_\mu C \,\bar{u}_n^T\right)\big{]}$ \\
  
$T_{cc\bar u\bar s}$&$\frac{1}{2}(0^{+})$
&$ {\cal O}_{T^{0^+}_{us}}   = \epsilon_{i j k} \:\epsilon_{m n k}
\left( c_i \, C \gamma_{\mu } c_j^T \right) 
\left( \bar{u}_m \,\gamma^{\mu } C \bar{s}_n^T \right)$\\

$T_{cc\bar s\bar s}$&$0(0^{+})$
&$ {\cal O}_T^{0^+}    = \epsilon_{i j k} \:\epsilon_{m n k}
\left( c_i \, C \gamma_{\mu } c_j^T \right) 
\left( \bar{s}_m \,\gamma^{\mu } C \bar{s}_n^T \right)$\\
   \\
   \hline\hline
  \vspace*{-0.5cm}
\end{tabular*}
}
 \caption{Interpolating operators describing the $ZXT$ states.}  

\label{tab:current}
\end{table*}
}
\section{Optimization criteria}
\nin As $\tau$, $t_c$ and $\mu$ are free external parameters, we shall use stability criteria (minimum sensitivity on the variation of these parameters) to extract the hadron masses and couplings.\\
\\
{\bf $\bullet$ $\tau$-stability:} The optimal result is obtained around a minimum or inflexion point. These optimal values of $\tau$ are equivalent to the so-called plateau used in the literature using the Borel $M^2\equiv 1/ \tau$ variable. At this $\tau$-stability region, one can check the lowest ground state dominance of the LSR and the convergence of the OPE.\\
{\bf $\bullet$ $t_c$-stability:} One expects it to be around the mass of the first radial excitation. To be conservative, we take $t_c$ from the beginning of $\tau$-stability until the $t_c$-stability starts to be reached. The $t_c$-stability region corresponds to a complete dominance of the lowest ground state in the QSSR.\\
{\bf $\bullet$ $\mu$-stability:} This is used to fix in a rigorous optimal way, the arbitrary subtraction constant appearing in the PT calculation of the Wilson coefficients and in the QCD input renormalized parameters. From our previous works, we have observed that the value of $\mu$-stability is (almost) universal for the different heavy-light four-quark states:
\beq
\mu_c=(4.65\pm 0.05)~{\rm GeV} ~~;~ \mu_b=(5.20\pm 0.05)~{\rm GeV}\, ,
\eeq
respectively for the charm and bottom channels. In the following analysis, we shall use these numbers.
\section{The $T_{cc\bar{u}\bar{d}}$ or $T_{cc}$ ($0^+$) state}
\nin We improve and extend the analysis in\,\cite{DRSR11,LEE,WANG-T,ZHU-T,MALT-T} using LSR and DRSR by including the FNLO PT contributions and by paying more carefully attention on the different sources of the errors.
\subsection*{$\lozenge$ Mass and decay constant from LSR at NLO}
\begin{figure}[H]

\begin{center}
\includegraphics[width=6.2cm]{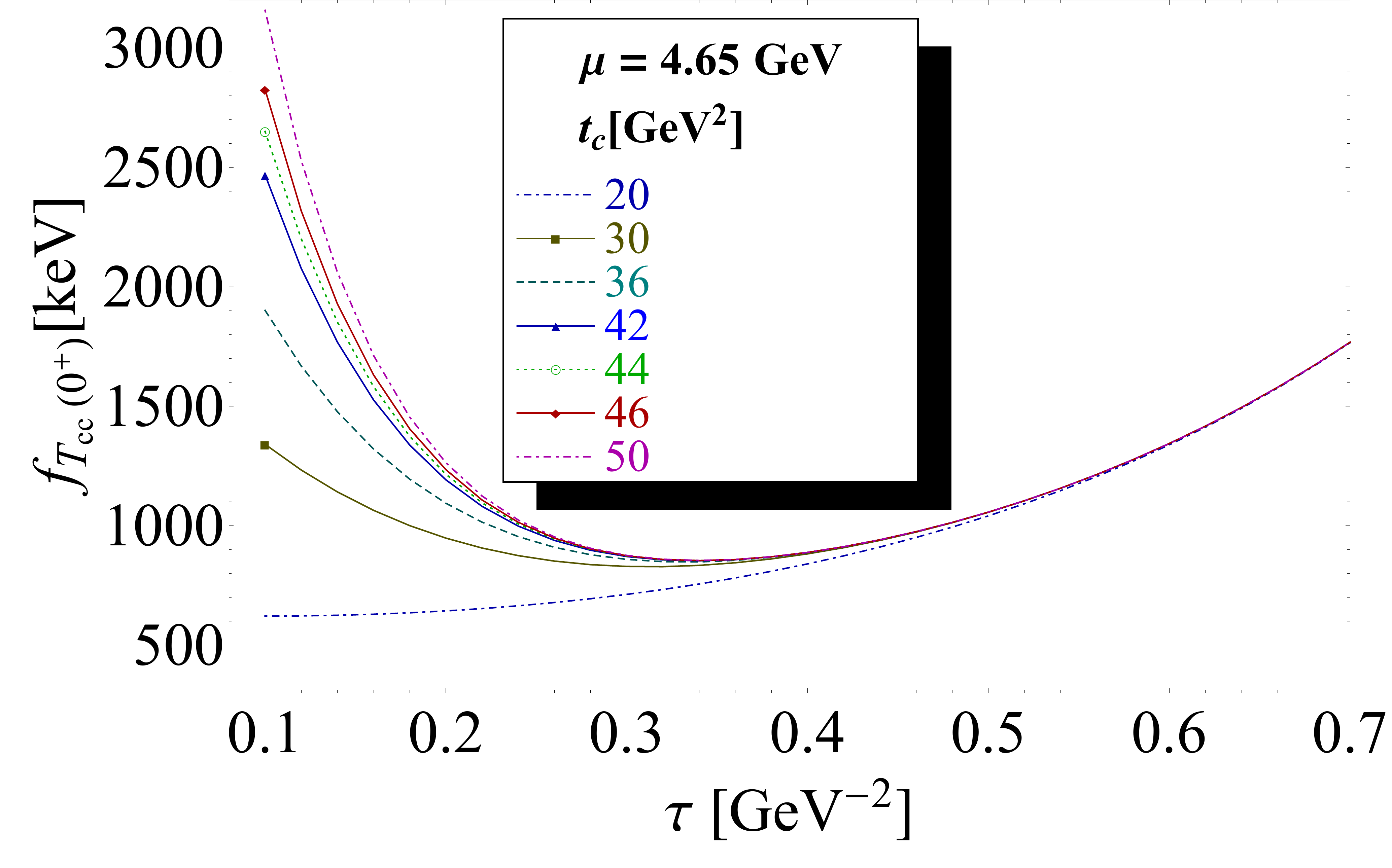}
\hspace*{0.5cm}
\includegraphics[width=6.5cm]{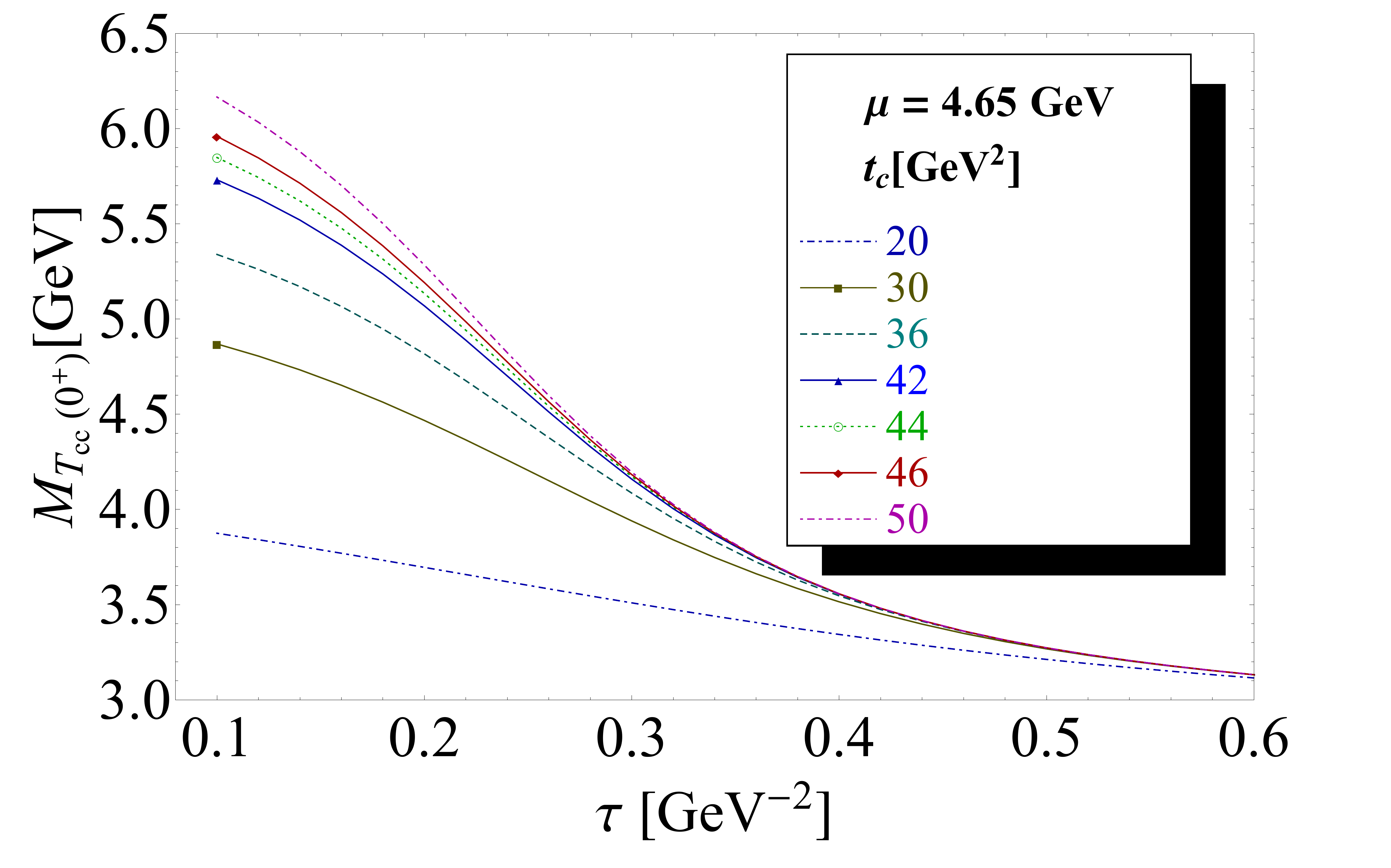}
\vspace*{-0.5cm}
\caption{\footnotesize  $f_{T_{cc}}(0^{+})$ and $M_{T_{cc}}(0^{+})$ as a function of $\tau$ at NLO for \# values of $t_c$ and for $\mu$=4.65 GeV.} 
\label{fig:tcc0p}
\end{center}
\vspace*{-0.5cm}
\end{figure}

\nin
We study the behavior of the coupling and mass in term of the LSR variable $\tau$ for different values of $t_c$ at NLO as shown in Fig.\,\ref{fig:tcc0p}. The optimal results are obtained with the sets\,:$(\tau,t_c)$=(0.31,30) to (0.34,46) (GeV$^{-2}$,GeV$^2$):
\beq
\hspace*{-0.12cm} f_{T_{cc}}(0^{+})\hspace*{-0.1cm} = \hspace*{-0.1cm} 841(83)~{\rm KeV}\, , M_{T_{cc}}(0^{+})\hspace*{-0.1cm}=\hspace*{-0.1cm} 3882(129)~{\rm MeV},
\label{eq:mt0cc}
\eeq
\subsection*{$\lozenge$ Ratio of masses $r_{T^{0^+}_{cc}/X_c}$ from DRSR}
\begin{figure}[hbt]
\begin{center}
\includegraphics[width=7cm]{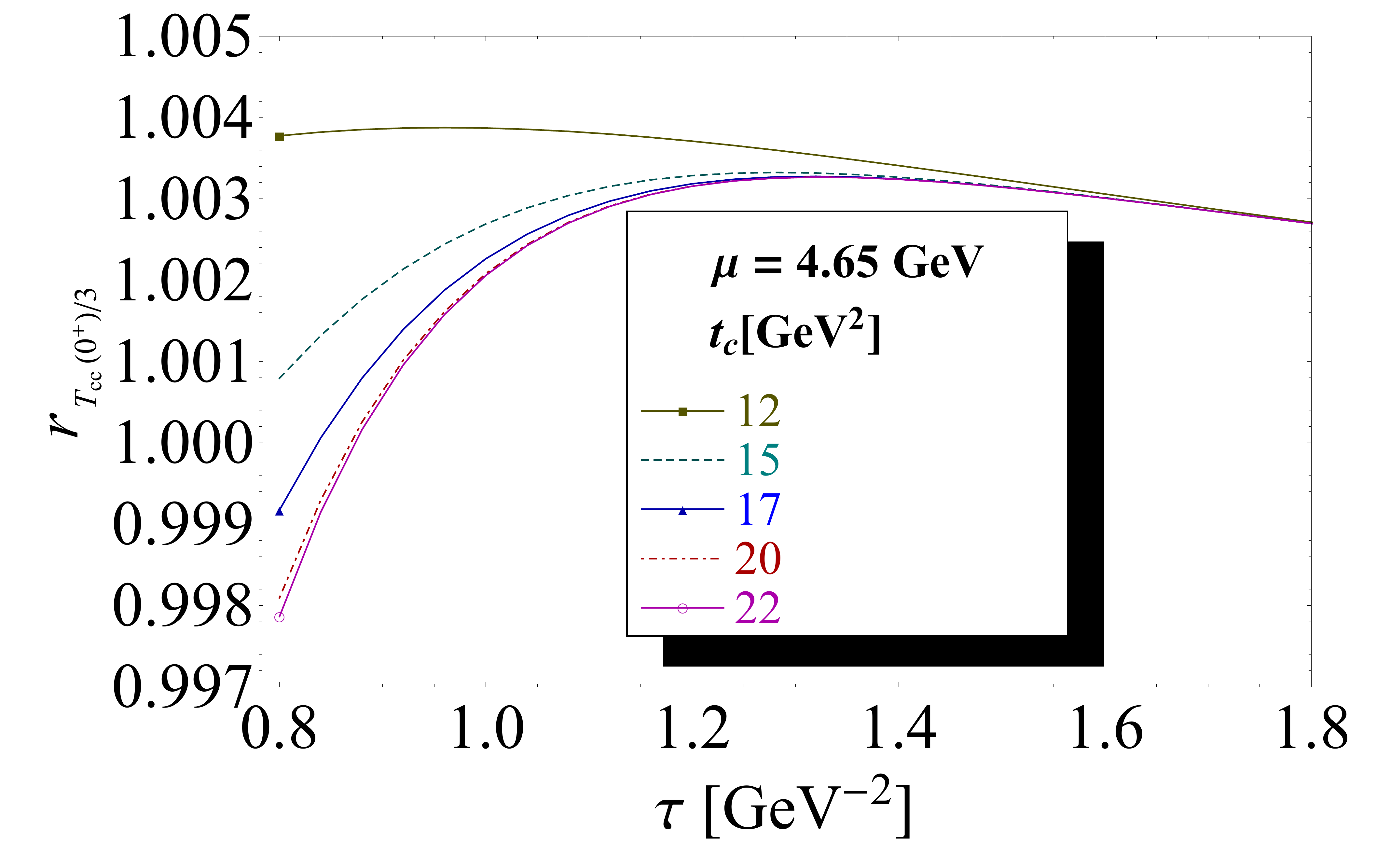}
\vspace*{-0.5cm}
\caption{\footnotesize   $r_{T^{0^+}_{cc}/X_c}$ as a function of $\tau$ at NLO for \# values of $t_c$ and for $\mu$=4.65 GeV.} 
\label{fig:rt0ccxc}
\end{center}
\end{figure}

\nin
We use the DRSR for studying the ratio of masses. The analysis is shown in Fig.\,\ref{fig:rt0ccxc}. The optimal result is obtained for the 
sets $(\tau,t_c)=$(1.28,15) to (1.32,20) $(\rm GeV^{-2},\rm GeV^2)$ from which one deduces:
\beq
r_{T^{0^+}_{cc}/X_c}= 1.0033(10)~~\lrar ~~ M_{T^{0^+}_{cc}}=3885(4)~{\rm MeV},
\eeq
Where we have used the experimental prediction for $X_c(3872)$\,\cite{X3872-EXP}.
\subsection*{$\lozenge$ Ratio of masses $r_{T^{0^+}_{cc}/T^{1^+}_{cc}}$ from DRSR}
\begin{figure}[hbt]
\begin{center}
\includegraphics[width=7cm]{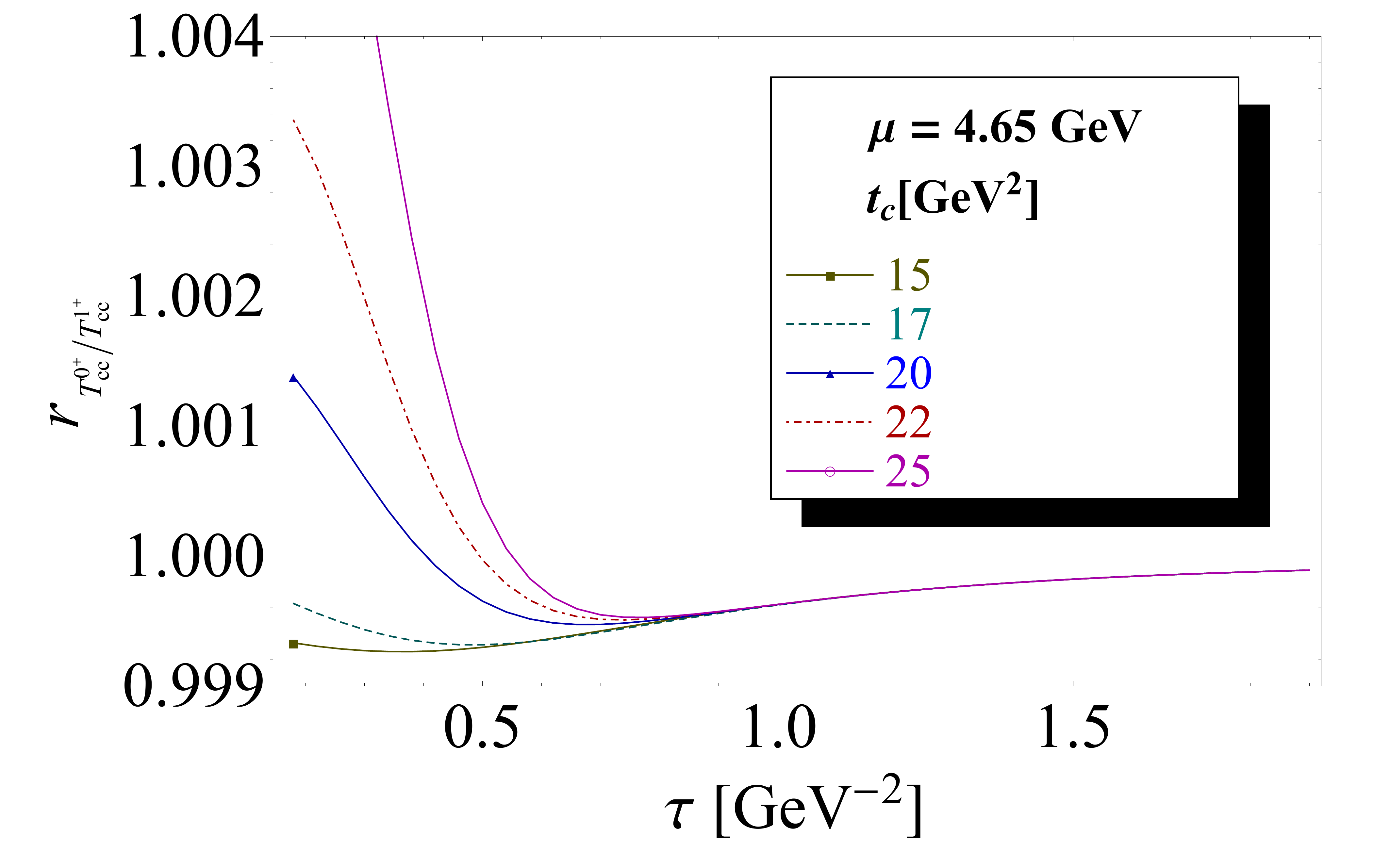}
\vspace*{-0.5cm}
\caption{\footnotesize   $r_{T^{0^+}_{cc}/T^{1^+}_{cc}}$ as a function of $\tau$ at NLO for \# values of $t_c$ and for $\mu$=4.65 GeV.} 
\label{fig:rtcc01p}
\end{center}
\end{figure}

\nin
We use the DRSR for studying the ratio of masses. The analysis is shown in Fig.\,\ref{fig:rtcc01p}. The optimal result is obtained for the sets $(\tau,t_c)=$(0.36,15) to (0.72,22) $(\rm GeV^{-2},\rm GeV^2)$ from which one deduces:
\beq
r_{T^{0^+}_{cc}/T^{1^+}_{cc}}= 0.9994(2)~~\lrar ~~ M_{T^{0^+}_{cc}}=3878(5)~{\rm MeV},
\eeq
Where we have used the mean from the predicted mass of $T_{cc}(1^+)$ in\,\cite{ANR22} and the data 3875 $\rm{MeV}$\,\cite{LHCb}.
\subsection*{$\lozenge$ Final value of $M_{T^{0^+}_{cc}}$ from LSR$\oplus$DRSR}
\nin
As a final value of $M_{T^{0^+}_{cc}}$, we take the mean of the previous three determinations:
\beq
M_{T^{0^+}_{cc}}=3883(3)~\rm{MeV}.
\eeq
\section{The $T_{cc\bar{s}\bar{u}}$ $0^+$ states}
\begin{figure}[hbt]
\vspace*{-0.25cm}
\begin{center}
\includegraphics[width=7cm]{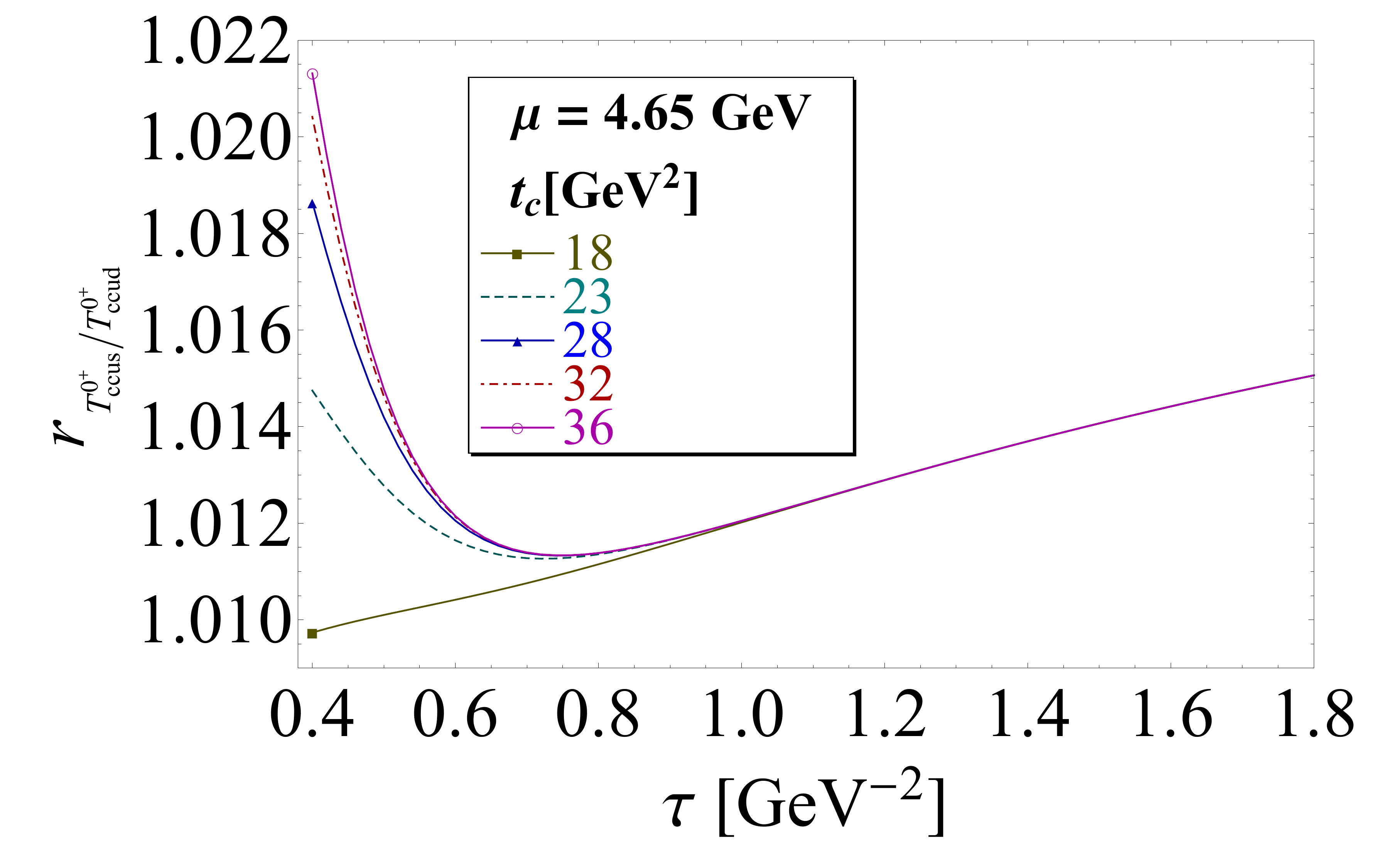}
\caption{\footnotesize   $r_{T^{0^+}_{cc\bar{s}\bar{u}}/T^{0^+}_{cc}}$ as a function of $\tau$ at NLO for \# values of $t_c$ and for $\mu$=4.65 GeV.} 
\label{fig:rtccs0p}
\end{center}
\end{figure}

\nin
The $\tau$ and $t_c$ behaviors of the coupling and mass from the LSR moments and their ratios are very similar to the ones in Fig.\ref{fig:tcc0p}. The optimal result is obtained for the 
sets $(\tau,t_c)=$(0.32,30) to (0.35,46) $(\rm GeV^{-2},\rm GeV^2)$ at which the coupling presents a minimum and the mass an inflexion point:
\beq
f_{T^{0^+}_{cc\bar{s}\bar{u}}} =  542(53)~{\rm KeV}~~~ ;~ M_{T^{0^+}_{cc\bar{s}\bar{u}}}= 3936(90)~{\rm MeV},
\label{eq:mt0ccs}
\eeq
For the SU(3) ratio of masses $r_{T^{0^+}_{cc\bar{s}\bar{u}}/T^{0^+}_{cc}}$, the analysis is shown in Fig.\ref{fig:rtccs0p}. The optimal result is obtained for the sets $(\tau,t_c)=$(0.72,23) to (0.74,32) $(\rm GeV^{-2},\rm GeV^2)$ at which we deduce:
\beq
\hspace*{-0.3cm} r_{T^{0^+}_{cc\bar{s}\bar{u}}/T^{0^+}_{cc}}= 1.0113(12)\lrar M_{T^{0^+}_{cc\bar{s}\bar{u}}}=3927(6)~{\rm MeV}.
\eeq
As a final result for the mass, we take the mean from the two determinations:
\beq
M_{T_{cc\bar{s}\bar{u}}}(0^+)=3927(6)~\rm{MeV}.\\
\eeq
\section{The $T_{cc\bar{s}\bar{s}}$ $0^+$ states}
\nin
The behaviors of the different curves (LSR and DRSR) are very similar to the ones in Fig.\ref{fig:tcc0p} and \ref{fig:rtccs0p}. The final result for the mass is given by the mean from the LSR at NLO and the ratios of masses $r_{T^{0^+}_{cc\bar{s}\bar{s}}/T^{0^+}_{cc}}$ from DRSR:
\beq
M_{T_{cc\bar{s}\bar{s}}}(0^+)=3993(11)~\rm{MeV}.
\eeq
\section{The $T_{bb\bar{q}\bar{q}'}$ $0^+$ states}
\nin
We complete our study by extending it to the bottom sector. Applying the same techniques and methods as in the charm channels, the final results are compiled in Table\,\ref{tab:b-results}.
\begin{table}[hbt]
\setlength{\tabcolsep}{1.3pc}
{\small
 {\begin{tabular}{@{}lrr@{}}
&\\
\hline
\hline
Obsevables& Couplings [$\keV$]& Masses [$\MeV$]\\
\hline
$T_{bb\bar{u}\bar{d}}$ & 54(11)&10484(63)\\
$T_{bb\bar{s}\bar{u}}$ &35(7)&10511(58)\\
$T_{bb\bar{s}\bar{s}}$ &47(13)&10567(57)\\
\hline\hline
\end{tabular}}
}
\caption{{\small $T_{bb\bar{u}\bar{d}}$ ($q,q'\equiv u,d,s$) couplings (LSR) and masses (LSR$\oplus$DRSR) predictions at NLO.}}
\label{tab:b-results}
\end{table}
\section{Summary}
\nin
-- Motivated by the recent LHCb discovery of a $1^{+}$ state at 3878 MeV\,\cite{LHCb}, just below the $D^*D$ threshold, which is a good isoscalar ($I=0$) $T_{cc\bar u\bar d}$ axial vector candidate, we extend our analysis for the scalar ($J^P=0^+$) $T_{ccqq'}$ states and improve the existing QSSR results by combining the direct mass determinations from the ratios ${\cal R}$ of LSR with the ratio of masses from DRSR within stability criteria. we have included to the LO perturbative term, the NLO corrections and contributions of quark and gluon condensates up to dimension-six. We have completed the analysis with new predictions of the future beauty $T_{bbqq'}$ states.\\
-- From our predictions, all the $T_{ccqq'}$ (resp. $T_{bbqq'}$) states are above (resp. below) their respective physical thresholds. One can also notice that the SU(3) breaking effects are positive and tiny.\\
-- Comparing our results with different QSSR predictions in the literature, one can conclude that within the errors, there are (almost) good agreement among different determinations. Most of the approaches predict the $T_{ccqq'}$ states to be above the hadronic thresholds while the $T_{bbqq'}$ is expected to be below the physical thresholds.\\
-- Confronted to different approaches in the literature (lattice calculations\,\cite{JUN,MALT,MOHAN,LESK},  light front holographic\,\cite{DOSCH2}, quark and potential models $\oplus$heavy quark symmetry\,\cite{MENG,ROSNER,QUIGG,BRAATEN,CHENG,RICHARD2,ZHUMODEL,WU,BICUDO}) one can also notice that our LSR$\oplus$DRSR predictions are (almost) in agreement (within the errors). The $T_{cc}$ states is expected to be above the physical threshold while the $T_{bb}$ ones are grouped around the threshold.\\
-- To sum up, our different predictions for $0^+$ states from LSR$\oplus$DRSR are clustered in the range $-250$ to $+150$ MeV of the physical thresholds.



\begin{thebibliography}{999}
\bibitem{ANR22} R. M. Albuquerque, S. Narison and D. Rabetiarivony, {\it Nucl. Phys.} {\bf A 1023}, 122451 (2022).
\bibitem{ANR21} R. M. Albuquerque, S. Narison and D. Rabetiarivony, {\it Phys. Rev.} {\bf D 103}, 074015 (2021); {\it Phys. Rev.} {\bf D 105}, 114035 (2022).
\bibitem{ANR1} R. Albuquerque et {\it al.}, {\it Nucl. Phys.} {\bf A 1007} (2021) 122113.
\bibitem{NR1} S. Narison and D. Rabetiarivony, {\it Nucl. Part. Phys. Proc.} {\bf 318-323}, (2022) 96-101.
\bibitem{ANRR2} R. M. Albuquerque et {\it al.}, {\it Nucl. Part. Phys. Proc.} {\bf 312-317}, (2021) 120-124.
\bibitem{QQQQ} R. M. Albuquerque et {\it al.}, {\it Phys. Rev.} {\bf D 102}, 094001 (2020).
\bibitem{ANRR1a} R. M. Albuquerque et {\it al.}, {\it Nucl. Part. Phys. Proc.} {\bf 300-302}, (2018) 186-195.
\bibitem{SU3} R. Albuquerque et {\it al.}, {\it Int. J. Mod. Phys.} {\bf A33} (2018), 1850082.
\bibitem{ANRR1} R. M. Albuquerque et {\it al.}, {\it Nucl. Part. Phys. Proc.} {\bf 282-284}, (2017) 83.
\bibitem{SNX1} R. Albuquerque et {\it al.}, {\it Int. J. Mod. Phys.} {\bf A31} (2016) no.17, 1650093.
\bibitem{SNX2} R. Albuquerque et {\it al.}, {\it Int. J. Mod. Phys.} {\bf A31} (2016) no. 36, 1650196.
\bibitem{AFNR} R. Albuquerque et {\it al.}, {\it Phys. Lett.} {\bf B 175}, (2012) 129.
\bibitem{ADKT} R. M. Albuquerque et {\it al.}, {\it J. Phys.} {\bf G 46}, 093002 (2019).
\bibitem{BELL} J.S. Bell and R.A. Bertlmann, {\it Nucl. Phys.} {\bf B177} (1981) 218.
\bibitem{BELLa} J.S. Bell and R.A. Bertlmann, {\it Nucl. Phys.} {\bf B187} (1981) 285.
\bibitem{BNR} C. Becchi et {\it al.}, {\it Z. Phys.} {\bf C 8}, 335 (1981).
\bibitem{BERT} R.A. Bertlmann, {\it Acta Phys. Austriaca} {\bf 53}, 305 (1981) and references therein.
\bibitem{NEUF}R.A. Bertlmann and H. Neufeld, {\it Z. Phys.} {\bf C 27} (1985)  437.
\bibitem{SNR}S. Narison and E. de Rafael,  {\it Phys. Lett.} {\bf B 522} (2001) 266.
\bibitem{SVZa} M.A. Shifman, A.I. Vainshtein and V.I. Zakharov, {\it Nucl. Phys.} {\bf B147} (1979) 385, 448.
\bibitem{Za} V.I. Zakharov, {\it Int. J. Mod. Phys.} {\bf A 14}, 4865 (1999).
\bibitem{SNB1} S. Narison, {\it QCD as a theory of hadrons, Cambridge Monogr. Part. Phys. Nucl. Phys. Cosmol.} {\bf 17} (2002) 1; [hep-ph/0205006].
\bibitem{SNB2}S. Narison, {\it QCD spectral sum rules ,  World Sci. Lect. Notes Phys.} {\bf 26} (1989) 1.
\bibitem{SNB3}S. Narison, {\it Phys. Rept.}  {\bf 84} (1982) 263; {\it Acta Phys. Pol.} {\bf B 26} (1995) 687. 
\bibitem{CK} E. de Rafael, hep-ph/9802448.
\bibitem{YND} F. J. Yndurain, {\it The Theory of Quark and Gluon Interactions,} 3rd ed. (Springer, New York, 1999).
\bibitem{PAS} P. Pascual and R. Tarrach, {\it QCD: Renormalization for Practitioner} (Springer, New York, 1985).
\bibitem{RRY} L. J. Reinders, H. Rubinstein, and S. Yazaki, {\it Phys. Rep.} 127, 1(1985).
\bibitem{IOFF} B. L. Ioffe, {\it Prog. Part. Nucl. Phys.} 56, 232(2006).
\bibitem{DOSCH} H. G. Dosch, {\it Non-Perturbative Methods}, edited by S. Narison (World Scientific, Singapor,1985).
\bibitem{X3872-EXP} P.A. Zyla et {\it al.} (Particle Data Group), {\it Prog. Theor. Exp. Phys.} {\bf 2020}, 083C01 (2020).
\bibitem{LHCb} R. Aaij et {\it al.} (LHCb Collaboration), {\it Nature Communications} {\bf 13}, (2022) 3351.
\\
\bibitem{DRSR88} S. Narison, {\it Phys. Lett.} {\bf B 210} (1988) 238.
\bibitem{PICH} A. Pich and E. de Rafael, {\it Phys. Lett.}  {\bf B158} (1985)  477.
\bibitem{NPIV} S. Narison and A. Pivovarov,  {\it Phys. Lett.} {\bf B 327} (1994) 341.
\bibitem{HAGIWARA}   K. Hagiwara, S. Narison and D. Namura, {\it Phys. Lett.}  {\bf B540} (2002) 233.
\bibitem{DRSR11}J. M. Dias, S. Narison, F.S. Navarra, M. Nielsen and J.-M. Richard, , {\it Phys. Lett.} {\bf B 703} (2011) 274.
\bibitem{LEE} F. S. Navarra, M. Nielsen and S. H. Lee, {\it Phys. Lett.} {\bf B 649} (2007) 166. 
\bibitem{WANG-T} Z.-G. Wang and Z. H. Yan, {\it Eur. Phys. J.} {\bf C 78} (2018 19; Z.-G. Wang, {\it Acta Phys. Pol.} {\bf 49} (2018) 1781.
\bibitem{ZHU-T} M.-L. Du, W. Chen, X.-L. Chen and S. L. Zhu, {\it Phys. Rev.} {\bf D 87} (2013) 014003.
\bibitem{MALT-T} L. Tang, B.-D. Wan, K. Maltman and C.-F. Qiao, {\it Phys. Rev.} {\bf D 101} (2020) 094032.
\bibitem{JUN} P. Junnarkar, N. Mathur, M. Padmanath,{\it Phys. Rev} {\bf D 99(3)} (2019) 034507.
\bibitem{MALT} A. Francis, R.J. Hudspith, R. Lewis, K. Maltman, {\it Phys. Rev. Lett.} {\bf 118(14)} (2017) 142001.
\bibitem{MOHAN} P. Mohanta and S. Basak, {\it Phys. Rev} {\bf D 102} (2020) 094516.
\bibitem{LESK} L. Leskovec, S. Meinel, M. Pflaumer and M. Wagner, {\it Phys. Rev} {\bf D 100} (2019) 014503.
\bibitem{DOSCH2} H.G. Dosch, S.J. Brodsky, G.F. de T\'eramond, M. Nielsen and L. Zou, {\it Nucl. Part. Phys. Proc.} {\bf 312-317} (2021) 135. 
\bibitem{MENG}Q.Meng, E.Hiyama, A.Hosaka, M.Oka, P.Gubler, K.U.Can, T.T.Takahashi and H.S.Zong,  {\it Phys. Lett.} {\bf B 814}(2021)136095.
\bibitem{ROSNER}M. Karliner, J.L. Rosner, {\it Phys. Rev} {\bf D 90 (9)} (2014) 094007.
\bibitem{QUIGG} E.J. Eichten, C. Quigg, {\it Phys. Rev. Lett.} {\bf 119 (20)} (2017) 202002.
\bibitem{BRAATEN}E. Braaten, L.-P. He and A. Mohapatra, {\it Phys. Rev. Lett.} {\bf D 103} (2021) 016001.
\bibitem{CHENG}J.-B. Cheng, S.-Y. Li, Y.-R. Liu, Z. G. Si, T. Yao,  {\it Chin. Phys.} {\bf C 45} (2021) 043102.
\bibitem{RICHARD2} E. Hernandez, J. Vijande, A. Valcarce and J.-M. Richard, {\it Phys. Lett.} {\bf B 800} (2020) 135073.
\bibitem{ZHUMODEL}X.-Z. Weng, W.-Z. Deng and S.-L. Zhu, arXiv: 2108.07242 [hep-ph]. 
\bibitem{WU}Y. Wu, X. Jin, R. Liu, H. Huang and J. Ping, arXiv: 2112.05967 (2021). 
\bibitem{BICUDO}P. Bicudo, K. Cichy, A. Peters, and M. Wagner, {\it Phys. Rev.} {\bf D 93} (2016) 034501.
\end{thebibliography}
\end{document}